\documentclass{sf2a-conf2012}
\usepackage{graphicx}

\usepackage[]{natbib}  
\usepackage[cyr]{aeguill}
\usepackage{epstopdf}

\def\BibTeX{{\rm B\kern-.05em{\sc i\kern-.025em b}\kern-.08em
    T\kern-.1667em\lower.7ex\hbox{E}\kern-.125emX}}
\bibpunct{(}{)}{;}{a}{}{,}  %%%%%%%%%%%%%  A&A bibliography style
\usepackage{bm}
\usepackage{amsmath} 
\usepackage{amsmath}
\usepackage{amssymb}

\newcommand{\m}[2]{m_{(#1),#2}}
\newcommand{\dm}[3]{m_{(#1),#2 #3}}   
\newcommand{\gab}[3]{g^{#1 #2}_{(#3)}}  
\def \zabi{z^{\alpha}_{A/B,i}}
\newcommand \dDdx[1]{\frac{\partial{\Delta}^{(1)}_{r}}{\partial x^{#1}} (\bm z(\lambda))}
\newcommand \dDdxab[1]{\frac{\partial{\Delta}^{(1)}_{r}}{\partial x_{A/B}^{#1}} (\bm z(\lambda))} 
\newcommand{\hti}[1]{\tilde h^i_{(#1)}}
\newcommand{\dhti}[2]{\tilde h^i_{(#1),#2}}
\newcommand{\dgab}[4]{g^{#1 #2}_{(#3),#4}}
\def\bx{{\bm x}}  
\def\ba{{\bm a}}  
\usepackage{hyperref}    
\begin{document}

\TitreGlobal{SF2A 2012}

%%-----------------------------------------------------------------
%%      the top matter
%%

\title{Frequency shift up to the 2-PM approximation}

\runningtitle{Frequency shift up to the 2-PM approximation}

\author{A. Hees$^{1,2,}$}\address{Royal Observatory of Belgium, Avenue Circulaire 3, 1180 Uccle, Belgium}
\address{SYRTE, Observatoire de Paris, CNRS, UPMC, Avenue de l'Observatoire 61, 75014 Paris, France}
\address{Namur Center for Complex Systems (naXys), University of Namur (FUNDP), Belgium}

\author{S. Bertone$^{2,}$}\address{INAF, Astronomical Observatory of Torino / University of Torino}

\author{C. Le Poncin-Lafitte$^2$}

%% Keep this line, even if the page will be settled afterwards.
\setcounter{page}{237}

%%-----------------------------------------------------------------

\maketitle

\begin{abstract}
%
% Warning!  within the abstract:
% - do not use macros. 
% - do not use commands like: \cite, \citet, \citep ... etc.
% 
A lot of fundamental tests of gravitational theories rely on highly precise measurements of the travel time and/or the frequency shift of electromagnetic signals propagating through the gravitational field of the Solar System. In practically all of the previous studies, the explicit expressions of such travel times and frequency shifts as predicted by various metric theories of gravity are derived from an integration of the null geodesic differential equations. However, the solution of the geodesic equations requires heavy calculations when one has to take into account the presence of mass multipoles in the gravitational field or the tidal effects due to the planetary motions, and the calculations become quite complicated in the post-post-Minkowskian approximation. This difficult task can be avoided using the time transfer function's formalism. 
We present here our last advances in the formulation of the one-way frequency shift using this formalism up to the post-post-Minkowskian approximation.
\end{abstract}

% Insert the keywords (to appear in the ADS indexing)
\begin{keywords}
frequency shift, relativity, fundamental physics, space navigation
\end{keywords}

\section{Introduction}
%---------------------

\noindent The treatment of light propagation in a relativistic framework is extremely important for various fields of study such as fundamental physics and astronomy, astrophysics and space navigation. Attaining very accurate measurements could allow us to observe a new range of subtle physical effects. Nowadays, a few approaches exist to model light propagation in a relativistic context. Among them, the post-Newtonian (pN) and the post-Minkowskian (pM) approximations~\citep[see for example][]{1999PhRvD..60l4002K,2010CQGra..27g5015K} are those mainly used in order to find perturbative solutions of the null geodesic equation. 

\noindent In this work, an alternative formulation to compute the one way frequency shift of an electromagnetic signal is presented. Being based on the time transfer function formalism~\citep{2008CQGra..25n5020T}, it allows us to compute this observable up to the post-post-Minkowskian approximation without integrating the null geodesic equation, allowing lighter calculations. 

\noindent Section~\ref{sec:not} contains the notations and conventions used in this document. 
In section~\ref{sec:1wayfreqshift} we give our framework and the definition of the one way frequency shift while in section~\ref{sec:TTF} we provide the relations between the frequency shift and the time transfer function.
These quantities will be then used in section~\ref{sec:comput} to obtain the post-Minkowskian expansion of the observable up to the 2PM approximation.
Our conclusions and possible applications of this study are given in section~\ref{sec:concl}.

\section{Notation and conventions}  \label{sec:not}
In this paper $c$ is the speed of light in a vacuum and $G$ is the Newtonian gravitational constant. The Lorentzian metric of space-time $V_4$ is denoted by $g$. The signature adopted for $g$ is $(+---)$. We suppose that space-time is covered by some global quasi-Galilean coordinate system $(x^\mu)=(x^0,\bx )$, where $x^0=ct$, $t$ being a time coordinate, and $\bx=(x^i)$. We assume that the curves of equations $x^i$ = const are timelike, which means that $g_{00}>0$ anywhere. We employ the vector notation $\ba$ in order to denote $(a^1,a^2,a^3)=(a^i)$. 
%Considering two such quantities $\ba$ and $\bb$ we use $\ba \cdot \bb$ to denote $a^ib^i$ (Einstein convention on repeated indices is used). The quantity $\vert \ba \vert$ stands for the ordinary Euclidean norm of $\ba$. 
For any quantity $f(x^{\lambda})$, $f_{, \alpha}$ denotes the partial derivative of $f$ with respect to $x^{\alpha}$. The indices in parentheses characterize the order of perturbation. They are set up or down, depending on the convenience. 

\section{The one-way frequency shift} \label{sec:1wayfreqshift}
%-------------------------
Consider a clock ${\cal O}_{\cal A}$ located at point ${\cal A}$ and a clock ${\cal O}_{\cal B}$ located at point ${\cal B}$ delivering, respectively, the proper frequency $\nu_{\cal A}$ and $\nu_{\cal B}$. Then, suppose that ${\cal O}_{\cal A}$ is sending an electromagnetic signal to ${\cal O}_{\cal B}$ along null geodesics of the metric (geometric optics approximation). Then, the one way frequency shift is defined by
\begin{equation}
\label{FreqOneWay}
\left . \frac{\Delta \nu}{\nu}\right \vert_\textrm{A$\rightarrow$ B}^\textrm{one-way}= \frac{\nu_B}{\nu_A} - 1\, .
\end{equation}

\noindent It is well-known that the ratio $\nu_B/\nu_A$ can be expressed as \citep{synge1960relativity}
\begin{equation}     \label{nubnua}
\frac{\nu_B}{\nu_A}=\frac{u_B^\mu k^{B}_\nu}{u_A^\nu k^{A}_\mu} 
=\frac{k^{B}_0}{k^{A}_0} \frac{u_B^0 + u_B^i \hat{k}^{B}_i}{u_A^0 + u_A^i \hat{k}^{A}_i}= \left( \frac{d \tau}{d t} \right)_A \frac{d t_A}{d t_B} \left( \frac{d t}{d \tau} \right)_B \, ,
\end{equation}
where  $u_{A/B}^\mu=(dx^\mu/ds)_{A/B}$ are the four-velocity of observers $\cal A$ and $\cal B$, $\hat{k}_i = \left(\frac{k_{i}}{k_{0}}\right)$ and $k^{A}_\mu$ and $k^{B}_\mu$ are the wave vectors (the null tangent vectors) at the point of emission $x_A$ and at the point of reception $x_B$, respectively.
Terms appearing in the right hand side of Eq. (\ref{nubnua}) can be expressed as
%$$\left( \frac{d \tau}{d t} \right)_{A/B} = \left[ g_{00}+2g_{0i} \beta^i+g_{ij} \beta^i\beta^j \right]^{1/2}_{A/B} \, ,$$  
\begin{equation}
\left( \frac{d \tau}{d t} \right)_{A/B} = \left[ g_{00}+2g_{0i} \beta^i+g_{ij} \beta^i\beta^j \right]^{1/2}_{A/B} \, ,\quad\frac{d t_A}{d t_B}=\frac{k^{B}_0}{k^{A}_0} \frac{1 + \beta_B^i \hat{k}^{B}_i}{1 + \beta_A^i \hat{k}^{A}_i}	\, ,
\end{equation} 
with $\beta^i_{A/B}=\frac{1}{c}\frac{dx^i_{A/B}}{dt}$ being the coordinate velocities of observers ${\cal A}$ and ${\cal B}$.
                      
\section{Relation between frequency shift and time transfer functions}  \label{sec:TTF}
We put $x_A = (ct_A, \bx_A)$ the event of emission ${\cal A}$ and $x_B = (ct_B, \bx_B)$ the event of reception ${\cal B}$. Moreover, we define ${\cal T}_{e}$ and ${\cal T}_{r}$ as two distinct (coordinate) time transfer functions defined as  
\begin{equation}  \label{ttf}
t_B - t_A = {\mathcal T}_e(t_A, \bx_A, \bx_B) = {\mathcal T}_r(t_B, \bx_A, \bx_B) \, .
\end{equation}
The relations between time transfer functions and the wave vectors $k^{\mu}=dx^{\mu}/d\lambda$ at emission and reception has been derived by \citet{2004CQGra..21.4463L} :
\begin{subequations}
\begin{eqnarray}  
& &\left(\widehat{k}_i\right)_A = \left(\frac{k_i}{k_0}\right)_A =
c \, \frac{\partial  {\cal T}_{e}}{\partial x^{i}_{A}}
\left[1 + \frac{\partial  {\cal T}_{e}}{\partial t_{A}}
\right]^{-1} \, = \, c \, \frac{\partial {\cal T}_{r}}{\partial x^{i}_{A}} \, , \label{2d2} \\
& & \nonumber \\
& &\left(\widehat{k}_i\right)_B = \left(\frac{k_i}{k_0}\right)_B =
-c \, \frac{\partial {\cal T}_{e}}{\partial x^{i}_{B}} \, = \,
- c \, \frac{\partial  {\cal T}_{r}}{\partial x^{i}_{B}}
\left[1 - \frac{\partial  {\cal T}_{r}} {\partial t_B}\right]^{-1}\,  , \label{2d1} \\
& & \nonumber \\
& &\frac{(k_{0})_B}{(k_{0})_A} =  \left[1 + 
\frac{\partial  {\cal T}_{e}}{\partial t_{A}}\right]^{-1} \, 
= \, 1 -
\frac{\partial  {\cal T}_{r}}{\partial t_{B}}   \, , \label{2d3}
\end{eqnarray} 
\end{subequations}
\noindent   
where $\mathcal T_e$ and $\mathcal T_r$ are evaluated at the event of emission ${\cal A}$ and at  the event of reception ${\cal B}$ respectively. It's then straightforward to define the one-way frequency shift~(\ref{FreqOneWay}) as a function of ${\cal T}_{e/r}$ and their partial derivatives.

\section{Post-Minkowskian expansion of the frequency shift} \label{sec:comput}
The expression of the time transfer functions $\mathcal{T}_{e/r}$ as a formal post-Minkowskian series has been derived by \citet{2008CQGra..25n5020T}. In this communication, we focus on ${\cal T}_{r}$, but similar considerations hold for ${ \cal T}_{e}$. ${ \cal T}_{r}$ can be written in ascending power of $G$ defined as
\begin{equation}
{\cal T}_{	r}(\bx_A, t_B, \bx_B)=\frac{R_{AB}}{c}+ \frac{1}{c}\sum_{n=1}^\infty\Delta^{(n)}_{r}(\bx_A, t_B,\bx_B)\, ,
\end{equation}
where $\Delta^{(n)}_{r}$ is of the order $\mathcal{O}(G^n)$, $R^i_{AB}=x^i_B-x^i_A$, $R_{AB}=\vert {R^i_{AB}} \vert$ and $N^i=\frac{R_{AB}^i}{R_{AB}}$.

\noindent Then, we can express the one-way frequency shift~(\ref{nubnua}) as follows (in agreement with the expression found by~\citet{2012arXiv1201.5041H})
\begin{equation}         \label{main}
\frac{\nu_B}{\nu_A}= \frac{\left[ g_{00}+2g_{0i} \beta^i +g_{ij} \beta^i\beta^j \right]^{1/2}_A}{\left[ g_{00}+2g_{0i} \beta^i +g_{ij} \beta^i\beta^j \right]^{1/2}_B}\times	 \frac{1-N^i \beta^i_B -\beta^i_B \frac{\partial \Delta_r}{\partial x^i_B}-\frac{\partial \Delta_r}{\partial t_B}}{1-N^i \beta^i_A + \beta^i_A \frac{\partial \Delta_r}{\partial x^i_A}}.
\end{equation} 
The goal of this work is to provide a new way of computing a general form for the derivatives of the time delay function up to the post-post Minkowskian order. In order to do so, we rewrite $\Delta_r^{(1)}$ and $\Delta_r^{(2)}$ given by~\citet{2008CQGra..25n5020T} as
\begin{subequations}
\begin{eqnarray}
	\Delta_r^{(1)} (\bx_A,t_B,\bx_B)&=& \frac{R_{AB}}{2} \int_0^1  \left[g^{00}_{(1)} - 2N^i g^{0i}_{(1)} + N^i N^j g^{ij}_{(1)}  \right]_{z^\alpha(\lambda)} d\lambda =  \int_0^1 m_{(1)}(\lambda) d\lambda \label{eq:Dr1PM}  \, ,  \\
	\Delta_r^{(2)}(\bx_A,t_B,\bx_B) &=& \int_0^1 \left[ \mathcal I_1(\lambda) + \mathcal I_2 (\lambda) +\mathcal I_3(\lambda) \right] d\lambda\, ,
\end{eqnarray}      
\end{subequations}
where   
\begin{subequations}
\begin{eqnarray}
	\mathcal I_1 &=& m_{(2)} (\lambda) - \Delta_r^{(1)} (\bm z(\lambda),t_B,\bm x_B)\  \m10(\lambda) \, , \\ 
	\mathcal I_2 &=& \left[ R_{AB}\  \gab 0i1 - R^k_{AB}\  \gab ik1  \right]_{z^\alpha(\lambda)}  \dDdx i \, ,  \\
	\mathcal I_3 &=& - \frac{R_{AB}}{2} \sum_{j=1}^3 \left[ \dDdx{j} \right]^2\, ,
\end{eqnarray}      
\end{subequations}
\noindent $\bm z(\lambda)=x^i_B-\lambda R^i_{AB}$ being the spatial composantes of the Minkowskian straight line of equation $z^\alpha (\lambda)=(x^0_B-\lambda R_{AB},\bm z(\lambda))$ and 
\begin{subequations}
\begin{eqnarray}
m_{(n),\alpha}(\lambda) &=& \frac{R_{AB}}{2} \left[g^{00}_{(n),\alpha} - 2N^i g^{0i}_{(n),\alpha} + N^i N^j g^{ij}_{(n),\alpha}  \right]_{z^\beta(\lambda)}\, ,  \\    \dDdx i &=& \int_0^1 \left[ m_{(1),\alpha}(\lambda\mu) z^\alpha_{A,i}(\lambda\mu) + \tilde h^i_{(1)}(\lambda\mu) \right] d\mu  \, , \nonumber
\end{eqnarray} 
\end{subequations}                                                                                                               
with $z^\alpha_{A/B,i}=\partial z^\alpha / \partial x^i_{A/B}$.
The derivatives of the first PM order of the delay function can be then easily calculated
\begin{subequations} \label{dDelta1} 
\begin{eqnarray}
	\frac{\partial{\Delta}^{(1)}_{r}}{\partial x^{i}_{A/B}} (\bx_A, t_B, \bx_B) &=& \int_0^1  \left[ m_{(1),\alpha}(\lambda) z^\alpha_{A/B,i}(\lambda) \pm \tilde h^i_{(1)} (\lambda) \right] d\lambda\, , \\
	\frac{\partial{\Delta}^{(1)}_{r}}{\partial t_{B}} (\bx_A, t_B, \bx_B) &=& \int_0^1  \left[ m_{(1),0}(\lambda) \; c \right] d\lambda\, , 
\end{eqnarray} 
\end{subequations}    
where the function $\tilde h$ is defined by
\begin{equation}
		\tilde h^i_{(n),j}(\lambda)= \left.\frac{\partial m_{(n),j}}{\partial x^i_A}\right|_{z^\alpha={\rm cst}}=-\left.\frac{\partial m_{(n),j}}{\partial x^i_B}  \right|_{z^\alpha={\rm cst}} = \frac{1}{2}\left[-N^ig^{00}_{(n),j}+2g^{0i}_{(n),j}-2g^{ik}_{(n),j}N^k+N^kN^lN^i g^{kl}_{(n),j}\right]_{z^\alpha(\lambda)}.  
\end{equation}
These equations are equivalent to those derived by \citet{2012arXiv1201.5041H}.
The same  approach can be used for the 2PM order resulting in more complex formulas                                                                               
\begin{subequations}   \label{dDelta2} 
\begin{eqnarray}
	\frac{\partial{\Delta}^{(2)}_{r}}{\partial x^{i}_{A/B}} (\bx_A, t_B, \bx_B) &=& \int_0^1  \left[\frac{\partial \mathcal I_1}{\partial x^i_{A/B}} +\frac{\partial \mathcal I_2}{\partial x^i_{A/B}}+\frac{\partial \mathcal I_3}{\partial x^i_{A/B}}\right] d\lambda \\
\frac{\partial{\Delta}^{(2)}_{r}}{\partial t_B} (\bx_A, t_B, \bx_B) &=& \int_0^1  \left[\frac{\partial \mathcal I_1}{\partial t_B} +\frac{\partial \mathcal I_2}{\partial t_B}+\frac{\partial \mathcal I_3}{\partial t_B}\right] d\lambda 
\end{eqnarray}    
\end{subequations}
where the derivatives can be expressed as follows  
\begin{subequations}\label{k2PM}
\begin{eqnarray} 
	\frac{\partial \mathcal I_1}{\partial x^{i}_{A/B}} &=&  \m2\alpha \zabi \pm \hti2 - \Delta_r^{(1)}(\bm{z}(\lambda),t_b, \bm x_B) \left[ \dm10\alpha \zabi \pm \ \dhti10  \right] -  \m10 \dDdxab{i}\, , \nonumber\\
	&& \\ 
\frac{\partial \mathcal I_2}{\partial x^{i}_{A/B}} &=& \left[ \mp N^i \gab0j1 \pm \gab ij1 +(R_{AB} \dgab 0j1\alpha  - \dgab jk1\alpha R^k_{AB}) \zabi \right] \dDdx{j}\\
&& \quad + [ R_{AB} \gab 0j1 -  R^k_{AB} \gab jk1 ] \frac{\partial^2{\Delta}^{(1)}_{r}}{\partial x^{i}_{A/B}\partial x^{j}} (\bm z(\lambda)) \,,\nonumber\\
  \frac{\partial \mathcal I_3}{\partial x^{i}_{A/B}} &=& \pm \frac{N_{AB}^i}{2} \sum_{j=1}^3 {\left(\dDdx{j}\right)^2}- R_{AB} \sum_{j=1}^3 \left[ \dDdx{j}  \cdot \frac{\partial^2{\Delta}^{(1)}_{r}}{\partial x^{i}_{A/B}\partial x^{j}} (\bm z(\lambda) ) \right] \, ,
%&&\rquad	\frac{\partial \mathcal I_1}{\partial t_B} =c \m20  - c \dm100  \Delta_r^{(1)}(z(\lambda)) -  \m10 \frac{\partial \Delta_R^{(1)}}{\partial t_B}(\bm z(\lambda)) \nonumber \\
%&&\rquad	\frac{\partial \mathcal I_2}{\partial t_B} =c [ R_{AB} \dgab 0i10 - R^k_{AB} \dgab ik10] \cdot \dDdx i + [ R_{AB} \gab 0i1 - R^k_{AB} \gab ik1] \cdot \frac{\partial^2{\Delta}^{(1)}_{r}}{\partial t_{B}\partial x^{i}} (z(\lambda)) \, ,\nonumber \\
%&&\rquad	\frac{\partial \mathcal I_3}{\partial t_B} = -R_{AB} \sum_{j=1}^3 {\dDdx{j} \cdot \frac{\partial^2{\Delta}^{(1)}_{r}}{\partial x^{i}_{A/B}\partial x^{j}} (z(\lambda))} \left.  \right\} d\lambda \; .
\end{eqnarray}
\end{subequations}          
all quantities being taken at $(\lambda)$ and where we define
\begin{equation}
	\frac{\partial^2{\Delta}^{(1)}_{r}}{\partial x^{i}_{A/B}\partial x^{j}} (\bm z(\lambda))=\int_0^1 \left[ \right. m_{(1),\alpha \beta} z^\alpha_{A,j} z^\beta_{A/B,i} \pm \tilde h^i_{(1),\alpha} z^\alpha_{A,j} + m_{(1),\alpha} z^\alpha_{AA/AB,ji} +\tilde h^j_{(1),\alpha} z^\alpha_{A/B,i} \pm \bar h^{ji}_{(1)} \left. \right]_{\lambda \mu} d\mu \, ,
\end{equation}
with the function $\bar h$ defined by
	\begin{equation}
	   	\bar h^{ik}_{(n)}=\left.\frac{\partial \tilde h^i_{(n)}}{\partial x^k_A}  \right|_{z^\alpha={\rm cst}}=-\left.\frac{\partial \tilde h^i_{(n)}}{\partial x^k_B}  \right|_{z^\alpha={\rm cst}} \, ,
	 \end{equation}
\noindent and $z^\alpha_{AA/AB,ji}=\frac{\partial^2 z^\alpha}{\partial x^j_A \partial x^i_{A/B}}$. Similar expressions can be written for $\partial \mathcal I_i/\partial t_B$.

\section{Conclusions} \label{sec:concl}
%--------------------
We presented here our last advances in the formulation of the one-way frequency shift up to the post-post-Minkowskian approximation. The main result is given by Eq. (\ref{main}) where the derivatives of $\Delta_r$ are given up to 2PM order by Eqs. (\ref{dDelta1}-\ref{dDelta2}). The advantage of our formulation is that it does not require the integration of the null geodesic differential equations. Instead, the frequency shift is expressed as integral of functions defined from the metric (and its derivatives) performed over a Minkowskian straight line. 

\noindent Exact formulas up to 2PM order may be required for future space missions exploring the inner Solar System as shown by~\citet{2010CeMDA.107..285T}. The formulas presented here are useful to derive the frequency shift directly from the space-time metric. They can be used to derive the frequency shift up to 2PM order in a Schwarzschild space-time to validate results from~\citet{2010CeMDA.107..285T}. One can also use them to derive the frequency shift in the field of an ensemble of moving point masses in PPN formalism or in a framework improving the current IAU conventions~\citep{2009PhRvD..79h4027M}. The present results can also be used to derive frequency shift in alternative theories of gravity if the corresponding space-time metric is known. These applications will be presented elsewhere in a forthcoming paper.

\begin{acknowledgements}
A. Hees is research fellow from FRS-FNRS (Belgian Fund for Scientific Research) and thanks FRS-FNRS for financial support.\\
S. Bertone is PhD student under the UIF/UFI (French-Italian University) program and thanks UIF/UFI for the financial support.
\end{acknowledgements}

%%-----------------------------
%%   Bibliography
%%-----------------------------
%
% The reference list should contain all the references cited in the text, ordered alphabetically by surname (with
% initials following). If there are several references to the same first author, they should be entered according
% to the following scheme:
% 1. One author: chronologically
% 2. Author, one co-author: alphabetically by co-author, then chronologically
% 3. Author, two or more co-authors: chronologically.
%
% Please note that for papers that have more than five authors, only the first three should be given, followed
% by "et al."
%
% The format for references is the one adopted by A&A (see the example below).
%
% To set the reference list in the proper A&A format, we encourage you to use BibTEX and the natbib
% package instead of the standard thebibliography environment.
%

\bibliographystyle{aa}  % A&A bibliography style file (aa.bst)
\bibliography{hees} % your references in file: Yourfile.bib

\begin{thebibliography}{9}
\expandafter\ifx\csname natexlab\endcsname\relax\def\natexlab#1{#1}\fi

\bibitem[{{Damour} \& {Esposito-Far{\`e}se}(1996)}]{1996PhRvD..53.5541D}
{Damour}, T. \& {Esposito-Far{\`e}se}, G. 1996, \prd, 53, 5541

\bibitem[{{Hees} {et~al.}(2012){Hees}, {Lamine}, {Reynaud}, {Jaekel}, {Le
  Poncin-Lafitte}, {Lainey}, {F{\"u}zfa}, {Courty}, {Dehant}, \&
  {Wolf}}]{2012arXiv1201.5041H}
{Hees}, A., {Lamine}, B., {Reynaud}, S., {et~al.} 2012, ArXiv e-prints

\bibitem[{{Klioner} \& {Zschocke}(2010)}]{2010CQGra..27g5015K}
{Klioner}, S.~A. \& {Zschocke}, S. 2010, Classical and Quantum Gravity, 27,
  075015

\bibitem[{{Kopeikin} \& {Sch{\"a}fer}(1999)}]{1999PhRvD..60l4002K}
{Kopeikin}, S.~M. \& {Sch{\"a}fer}, G. 1999, \prd, 60, 124002

\bibitem[{{Le Poncin-Lafitte} {et~al.}(2004){Le Poncin-Lafitte}, {Linet}, \&
  {Teyssandier}}]{2004CQGra..21.4463L}
{Le Poncin-Lafitte}, C., {Linet}, B., \& {Teyssandier}, P. 2004, Classical and
  Quantum Gravity, 21, 4463

\bibitem[{{Minazzoli} \& {Chauvineau}(2009)}]{2009PhRvD..79h4027M}
{Minazzoli}, O. \& {Chauvineau}, B. 2009, \prd, 79, 084027

\bibitem[{Synge(1960)}]{synge1960relativity}
Synge, J. 1960, Relativity: the general theory, Series in physics
  (North-Holland Pub. Co.)

\bibitem[{{Teyssandier} \& {Le Poncin-Lafitte}(2008)}]{2008CQGra..25n5020T}
{Teyssandier}, P. \& {Le Poncin-Lafitte}, C. 2008, Classical and Quantum
  Gravity, 25, 145020

\bibitem[{{Tommei} {et~al.}(2010){Tommei}, {Milani}, \&
  {Vokrouhlick{\'y}}}]{2010CeMDA.107..285T}
{Tommei}, G., {Milani}, A., \& {Vokrouhlick{\'y}}, D. 2010, Celestial Mechanics
  and Dynamical Astronomy, 107, 285

\end{thebibliography}

\end{document}